\documentclass[aps,pra,twocolumn,english,superscriptaddress,10pt]{revtex4-1}
\usepackage{verbatim}
\usepackage{amsmath}
\usepackage{amssymb}
\usepackage{graphicx}
\usepackage{braket}
\usepackage{xcolor,subfigure,upgreek,bbold}
\usepackage[colorlinks,linkcolor=blue,citecolor=blue,urlcolor=blue,breaklinks=true]{hyperref}
\usepackage{verbatim}
\usepackage{esint}

\graphicspath{{./figs/}{./}}

\begin{document}

\title{Resonant two-site tunnelling dynamics of bosons in a tilted optical superlattice}

\author{Anton S. Buyskikh}
\email{anton.buyskikh@strath.ac.uk}
\affiliation{Department of Physics and SUPA, University of Strathclyde, Glasgow G4 0NG, UK}

\author{Luca Tagliacozzo}
\affiliation{Department of Physics and SUPA, University of Strathclyde, Glasgow G4 0NG, UK}
\affiliation{Departament de F\'{\i}sica Qu\`antica i Astrof\'{\i}sica and Institut de Ci\`encies del Cosmos (ICCUB), Universitat de Barcelona,  Mart\'{\i} i Franqu\`es 1, 08028 Barcelona, Catalonia, Spain}

\author{Dirk Schuricht}
\affiliation{Institute for Theoretical Physics, Center for Extreme Matter and Emergent Phenomena, Utrecht University, Princetonplein 5, 3584 CE Utrecht, The Netherlands}

\author{Chris A. Hooley}
\affiliation{SUPA, School of Physics and Astronomy, University of St Andrews, North Haugh, St Andrews KY16 9SS, UK}

\author{David Pekker}
\affiliation{Department of Physics and Astronomy, University of Pittsburgh, Pittsburgh, PA 15260, USA}

\author{Andrew J. Daley}
\affiliation{Department of Physics and SUPA, University of Strathclyde, Glasgow G4 0NG, UK}

\date{\today}

\begin{abstract}
We study the non-equilibrium dynamics of a 1D Bose-Hubbard model in a gradient potential and a superlattice, beginning from a deep Mott insulator regime with an average filling of one particle per site.
Studying a quench that is near resonance to tunnelling of the particles over two lattice sites, we show how a spin model emerges consisting of two coupled Ising chains that are coupled by interaction terms in a staggered geometry.
We compare and contrast the behavior in this case with that in a previously studied case where the resonant tunnelling was over a single site.
Using optimized tensor network techniques to calculate finite temperature behavior of the model, as well as finite size scaling for the ground state, we conclude that the universality class of the phase transition for the coupled chains is that of a tricritical Ising point.
We also investigate the out-of-equilibrium dynamics after the quench in the vicinity of the resonance and compare dynamics with recent experiments realized without the superlattice geometry.
This model is directly realizable in current experiments, and reflects a new general way to realize spin models with ultracold atoms in optical lattices.
\end{abstract}

\maketitle

\section{Introduction}
\label{sec:intro}

In recent years, ultracold atoms in optical lattices have proven to be a flexible testing ground for phenomena in strongly interacting systems \cite{Lewenstein2012, Bloch2012}. These experimental platforms both allow us to implement and explore models that have been developed theoretically as models for complex condensed matter systems, and inspire us to consider physics that is motivated by these experiments and has no direct analogue in other physical systems. Particular interest lies in the time-dependent control of parameters, and especially immediate quantum quenches, leading to out-of-equilibrium dynamics that can be tracked in real time \cite{Cazalilla2010,Polkovnikov2011,Daley2014}.

Recent experimental work demonstrated many-body dynamics of bosons in a tilted optical lattice \cite{Simon2011,Meinert2013,Meinert2014} far away from regimes of a simple quantum walk of a single particle \cite{Thommen2004,Preiss2015}. Following a theoretical proposal by Sachdev et al., Ref.~\cite{Sachdev2002}, a regime of resonant tunnelling was explored where the energy to tunnel over one site in the lattice, $E$, is equal to the on-site energy shift between two atoms $U$. This system exhibits a quantum phase transition to a density-wave-ordered state, in which empty sites alternate with doubly occupied sites. This system also exhibits interesting and non-trivial many-body dynamics in out-of-equilibrium situations \cite{Rubbo2011,Meinert2013}.

The theory has been extended to higher dimensions \cite{Pielawa2011}, but has up to now been applied to one-site resonant tunnelling only. However, experiments have shown out-of-equilibrium resonant dynamics over multiple sites, when $E=U/n$ for positive integer $n$ \cite{Meinert2014}. Here and in Ref.~\cite{Buyskikh2018} we introduce a superlattice into this system, in order to cleanly extend these studies to the case of general $n$. We focus here on $n=2$, comparing and contrasting the critical behavior and the dynamics observed for $n=1$ and $n=2$. We derive effective spin models for each of the cases and analyze the dynamics of atoms from this perspective. In contrast to Ref.~\cite{Buyskikh2018}, here we analyse the behaviour of the structure factor of the spin model, as well as details of the collective excitations in the coupled spin chains. This is specifically relevant for measurements that could be performed in quantum gas microscopes. In addition, we present a general technique for the analysis of projects Hamiltonians with tensor network methods. 

This article is organized as follows: In Sec.~\ref{sec:model} we start by reviewing the effective skew-field Ising model describing resonant tunnelling over one site ($n=1$). We then derive the effective spin model in regime of two-site ($n=2$) resonant tunnelling, as arises in the superlattice geometry.
In Sec.~\ref{sec:eff_model} we consider elementary excitations in each model, comparing and contrasting the dynamics in the $n=1$ and $n=2$ cases. 
We then study more closely the critical behavior of each of the models in Sec.~\ref{sec:phase_transition}, making use of numerical calculations with finite size scaling at critical points.
We confirm the Ising criticality of the $n=1$ model and then determine that the scaling of the order parameter of the second model is compatible with a tricritical Ising point \cite{Guida1998}.
We conclude our investigation with comparison of the specific heat capacity in both cases in Sec.~\ref{sec:finite_temp}, and provide a summary and outlook in Sec.~\ref{sec:summary}.

\section{The spin models}
\label{sec:model}

In this section we derive effective spin models for motion of bosons on the optical lattice described by the one-dimensional Bose-Hubbard Hamiltonian
\begin{equation}
H=-J\sum_{\braket{i,j}} b_i^\dagger b_j+\frac{U}{2}\sum_i n_i (n_i-1)-\sum_i V_i n_i,\label{eq:H_BH}
\end{equation}
where $J$ is the hopping matrix element, $\braket{\cdot,\cdot}$ implies summation only over neighboring sites, $U$ is the on-site interaction between bosons, and the site-dependent external potential
\begin{equation}
V_i=E\cdot i+(-1)^i \frac{\mu}{2},
\end{equation}
which has two contributions.
The first is the external linear field $E$ that creates a constant gradient potential, which can be generated by either a gravitational force (if the optical lattice sites are oriented vertically) or an external field (electric or magnetic).
The second contribution to $V_i$ is the energy offset $\mu$ between even and odd sites of the optical lattice, defining the superlattice.
The model preserves the total number of particles $N$ and we will be considering the case of the unit filling, when the number of sites $M=N$.

In the presence of a linear tilt $E\neq0$ the energy of the Hamiltonian \eqref{eq:H_BH} for the infinite lattice is not bounded from below, hence the ground state cannot be properly defined.
It means that the stationary state is the one in which all particles have fallen down the potential ladder and left the system.

Instead we focus on the regime deep in the Mott insulator (MI) phase, i.e. $J\ll U$, and consider the dynamics of atoms on typical experimental timescales.
However, the system can be tuned to a situation in which there are very different time scales, so that starting from a certain initial state, for short enough time (this scale will be explained below for particular examples of tilts), the system will only visit certain states in the vicinity of the initial state, and only on much longer time scales the particles will eventually fall out of the lattice.

The experiment we have in mind thus would start with the unit filled MI state, $\prod_i \ket{1}_i$, and $V_i=0$, then the external potential is quenched on, $V_i \neq 0$, allowing particles to tunnel to other sites more effectively.
This experiment has already been realized in a number of laboratories \cite{Simon2011,Meinert2013,Meinert2014}, and the MI phase has been observed to be resilient for generic values of the linear tilt $E$ on the duration of the experiment, due to the strong interactions $U$.
Despite the tilt the bosons are trapped for long times at their initial positions, this phenomenon was expected, since as first discussed in \cite{Davies1988}, even without interactions a linear tilt supports localization of the energy eigenstates on single sites.

The situation changes for specific values of the linear tilt $E=U/n$, where $n$ is an integer number, as then the MI state resonantly couples to a subset of other states. The dynamics of bosons was observed experimentally in Refs.~\cite{Simon2011,Meinert2013,Meinert2014}.
The fact that only a small number of states is in resonance with the parent MI state suggests the existence of an effective model describing the behavior of bosons in the vicinity of the resonance.

The case of the nearest-neighboring resonant tunneling $E=U$ has been extensively studied in Refs.~\cite{Sachdev2002,Pielawa2011} in the case of one and two dimensions.
We revisit this case in Sec.~\ref{subsec:model_U1} as it plays a role as a building block for the case of the next-nearest-neighboring-site resonant tunneling, $E=U/2$, considered in Sec.~\ref{subsec:model_U2}.

\subsection{$E=U$ regime}
\label{subsec:model_U1}

\begin{figure}[tb]
\begin{centering}
\includegraphics[width=1\linewidth]{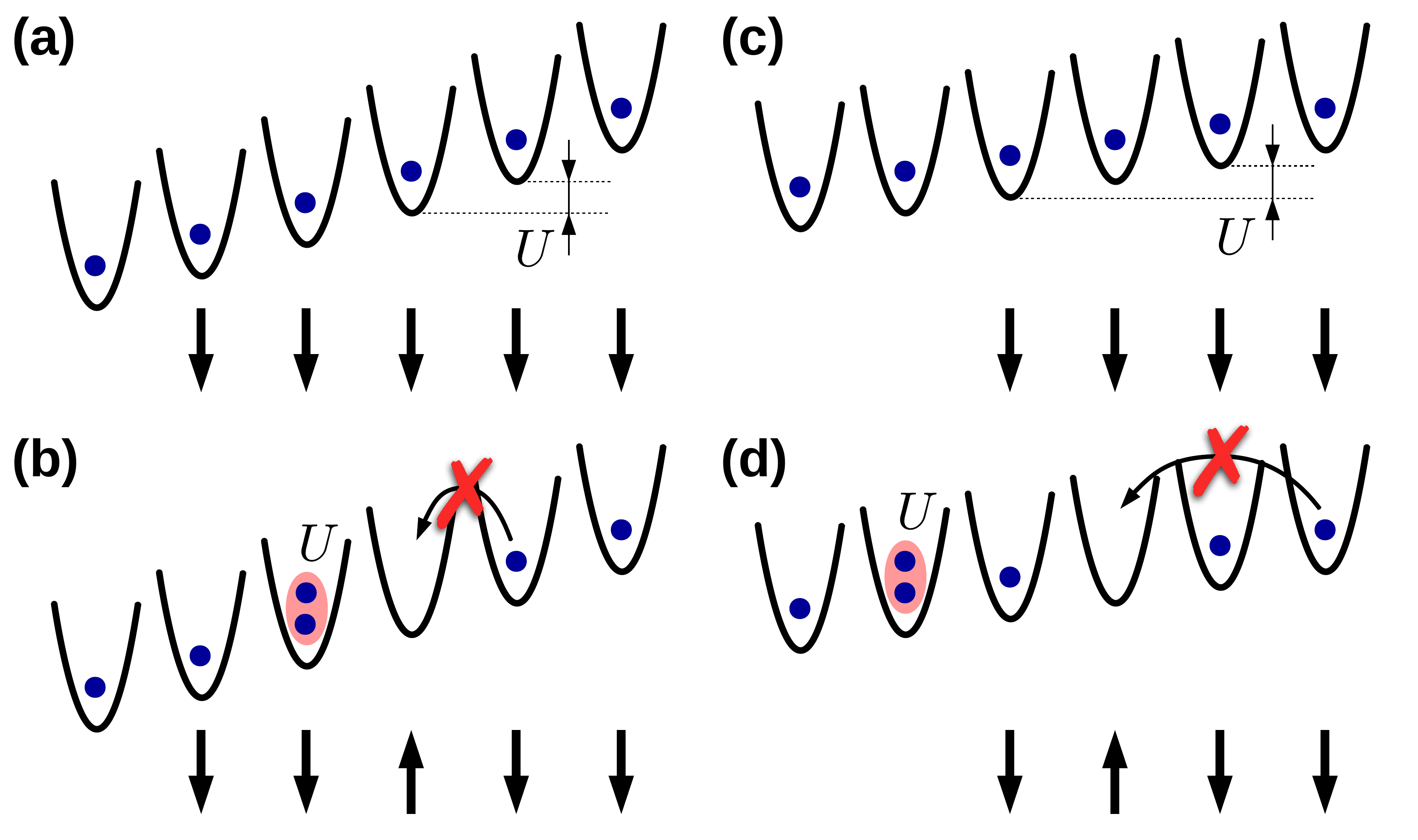}
\par\end{centering}
\caption{
\label{fig:illustration}
Schematic representation of the tilted unit filled MI states and example states coupled to them in the regime $E=U$ (a,b) and $E=U/2$ (c,d).
Mapping to spin chains works on the site-wise basis and drawn underneath.
For finite lattices, sites at the bottom of the tilted potential do not have mapping to spins because bosons on these sites cannot resonantly tunnel away.
In the regime $E=U$ the initial state $\prod_i\ket{1}_i$ is mapped to $\prod_i\ket{\downarrow}_i$ (a), if the $i^\mathrm{th}$ boson resonantly tunnels away the $i^\mathrm{th}$ spin changes to $\ket{\uparrow}$ (b).
Bosons from neighboring sites cannot hop down the slope simultaneously (b), which forbids configurations of two consecutive $\ket{\uparrow}$.
In the regime $E=U/2$ the situation is similar if a boson stays at the initial site, it is mapped to $\ket{\downarrow}$ (c), and if a boson tunnels \emph{two} sites down the slope it is mapped to $\ket{\uparrow}$ (d).
Bosons on next-nearest-neighboring sites cannot hop down the slope resonantly, which forbids two consecutive $\ket{\uparrow}$ on sites of the same parity (d).
The role of the superlattice offset $\mu$ (not shown here) is explained in the main text.}
\end{figure}

In the presence of the linear energy shift $E=U$ and $\mu=0$ the particles from the initial MI state (Fig.~\ref{fig:illustration}(a)) can hop to the nearest-neighboring site (down the tilted potential) lowering the potential energy by $E$.
Since the simultaneous gain of the interaction energy $U$ exactly compensates this energy change the two states become degenerate and it is said that they are resonantly connected (Fig.~\ref{fig:illustration}(b)).

Once a boson has tunneled resonantly from its initial site, the condition for resonant tunneling of its neighbors has changed and they cannot move freely.
In Fig.~\ref{fig:illustration}(b) one can see an example of this constraint, which in general means that states with two doubly occupied neighboring sites are not allowed.
In the regime $J\ll U$ transitions into those states are suppressed in perturbation theory by a factor $\propto J/U$, and hence the occupation of such states is suppressed as $\propto(J/U)^2$.
Exactly the same energy argument can be considered for other single-tunneling processes between the states, for which the energy difference is proportional to $U$.

A small detuning from the resonance condition, i.e. $|U-E| \ll U$, does not qualitatively alter the scenario and thus bosons can resonantly tunnel only between two sites: the initial one and its closest neighbor.
The system dynamics is then confined to the subspace of states resonantly connected by tunneling of the indistinguishable bosons to the parent tilted MI state.

This fact is exploited to establish the following mapping with a spin-$\frac{1}{2}$ chain: a boson on its initial position on the $i^\mathrm{th}$ site is mapped to a spin down $\ket{\downarrow}_i$, and a boson that leaves the $i^\mathrm{th}$ site via the resonant tunneling with a spin up $\ket{\uparrow}_i$.
Note that following this scheme, $\ket{\downarrow}_i$ can be associated with a site with one or two bosons on the $i^\mathrm{th}$ site of the lattice, but due to the constraint forbidding tunneling from neighboring sites this mapping is actually one-to-one and confusion is always avoided by checking the occupation of the neighboring sites (see Fig.~\ref{fig:illustration}(a,b)).

Then in the regime of a small detuning
\begin{equation}
|U-E|,J \ll E,U,
\end{equation}
the behavior of bosons at relevant time scales can be mapped to the effective spin model
\begin{equation}
H_U=\sum_i(-\sqrt{2}\sigma_i^x+\tilde\lambda\sigma_i^\uparrow+W\sigma_i^\uparrow\sigma_{i+1}^\uparrow),\label{eq:Heff_U1}
\end{equation}
where
\begin{equation}
\tilde\lambda=\frac{U-E}{J}
\end{equation}
denotes the deviation from the resonance, $\sigma^\uparrow=(\sigma^z+1)/2$ is a projector on spin-up, and $W\to+\infty$ is the constraint term that forbids two neighboring spin-ups. Note that the Hamiltonian \eqref{eq:Heff_U1} is dimensionless as it is rescaled by the first order tunneling amplitude $J$. It should be mentioned that a very similar Hamiltonian describes quantum dynamics of the ensemble of Rydberg atoms \cite{Bernien2017, Turner2018}, which was proposed back in \cite{Jaksch2000}.

\subsection{$E=U/2$ regime}
\label{subsec:model_U2}

Analogously to the regime $E=U$, we now build the effective spin model that describes the behavior of the unit filled MI state, but with a linear tilt $E=U/2$.
Besides this mapping will require a superlattice geometry $\mu\gg J$, which will confine bosons to sites of the same parity, i.e. bosons from initial odd sites will always move only to odd sites, and the same for bosons on even sites.
Hence the resulting effective model in this regime will resemble two spin chains $E=U$ that are coupled.

Once the tilt is set exactly at $E=U/2$ the initial MI state becomes degenerate with a set of other states, for instance in Fig.~\ref{fig:illustration}(c) one can see that if a boson moves two sites down the slope, the new state will be degenerate with the initial MI state.
The tunneling part of the Hamiltonian \eqref{eq:H_BH} will play the role of a perturbation that couples the states of this energy manifold.
The reader can see that in order to couple two resonant states at least two tunneling processes are required, which means that the construction of the effective Hamiltonian \cite{Bir1974} will require second order processes.

All the non-trivial resonant transitions in the energy manifold of interest can be categorized in three types.

The first type of transitions couple resonant states via tunneling of a boson over two sites down the slope of the tilted potential.
In this case the amplitude of the transition will depend on the occupations of the initial and final sites as well as the intermediate site.
For instance, the initial MI site is coupled with
\begin{equation}
\prod_i\ket{1}_i\leftrightarrow\ket{0}_j\ket{2}_{j+2}\prod_{i\neq j,j+2}\ket{1}_i,
\end{equation}
where a single boson tunnels twice ending up on a next-nearest-neighboring site.
This process can go via two channels: when the boson on site $j$ tunnels to $j+1$ and then from $j+1$ to $j+2$, or the boson from $j+1$ first tunnels to $j+2$ and then the boson from $j$ tunnels to $j+1$.
The resulting matrix element of this transition equals $3\sqrt{2}J^2/U$.
For each process of this type, there is the opposite, where a particle from a doubly occupied site tunnels uphill to the empty site.
Analogous to the regime $E=U$ one can notice that states with two doubly occupied sites on next-nearest-neighboring sites are not in the resonance manifold with the original MI state (Fig.~\ref{fig:illustration}(d)), i.e. the occupation of such states is suppressed as $\propto(J/U)^4$.

In the second type of resonant tunneling processes a single boson hops to its neighboring site and then hops back to the original site.
In this case the configuration of bosons does not change, but each state obtains an energy shift depending on the occupation of neighboring sites.
For instance, the initial MI state obtains the energy shift of $-16J^2/3U$ per boson, ignoring boundaries.

The third type of process occurs only for certain configurations of bosons on the lattice.
For these processes two bosons from the same site tunnel in the opposite directions, for instance the transition
\begin{eqnarray}
\ket{...,0,\overset{\curvearrowleft\curvearrowright}{1,2,0},1,2,...} & \leftrightarrow & \ket{...,0,2,0,1,1,2,...}
\end{eqnarray}
is resonant however it does not fit in the mapping scheme (will be explained below).

For this reason we introduce the offset energy $\mu\gg J$ between even and odd sites of the lattice, which means that the occupation of states achieved only via the third type of processes will scale
as $\propto(J/\mu)^4$.
For simplicity, we also assume from now on that $\mu\ll U$, which will make the effective spin model independent on $\mu$.
In principle, the flexible ratio $\mu/U$ can be exploited in the experiment, but we will leave this discussion for future work.

The reader might also think that there is another type of non-trivial processes when two bosons in different parts of the lattice tunnel to their neighboring sites in the opposite directions reaching another resonant state, e.g.
\begin{equation}
\ket{...,0,\overset{\curvearrowleft}{1,2},0,1,\overset{\curvearrowright}{2,1},...} \leftrightarrow \ket{...,0,2,1,0,1,1,2,...}
\end{equation}
However, these processes arises from two channels, depending whether it is the boson from the left or right site tunnels first.
Hence, these channels will have different intermediate states, more precisely the energy difference of these states and the energy of the MI state will have the same amplitude, but different sign.
Hence, the amplitudes of this channels added together cancel each other exactly.

The mapping between bosons and spins is similar to the regime $E=U$.
With only the first and second types of transition left, the dynamics of each boson is confined between the initial site of the parent MI state and the resonantly connected site.
That is why it is enough to consider mapping to the spins-$\frac{1}{2}$ chain, for each particle located on its initial $i^\mathrm{th}$ site of the MI state we assign a spin down $\ket{\downarrow}_i$ (Fig.~\ref{fig:illustration}(c)).
Then if a boson from the site $i$ tunnels to the next-nearest-neighboring site down the slope the corresponding spin becomes $\ket{\uparrow}_i$ (Fig.~\ref{fig:illustration}(d)).
Note that in the case of open boundary conditions the bosons on the last two sites can resonantly tunnel only via the second type of process, but not the first one.
That is why their corresponding spins states are always $\ket{\downarrow}$ and they can be eliminated from the spin model.

With this mapping scheme, the first type of process provides the tunneling part of the effective spin Hamiltonian
\begin{equation}
H_{U/2}^\mathrm{tun}=\sqrt{2}\frac{J^2}{U/2}\sum_i(\sigma_i^x+2\sigma_{i-1}^z\sigma_i^x+2\sigma_i^x\sigma_{i+1}^z),\label{eq:Heff_U2_tun}
\end{equation}
where we have a new characteristic energy $J^2/(U/2)$ instead of $J$ for $E=U$.
Another interesting thing is that besides the first term corresponding to a simple spin flipping, one can see additional terms that modify the amplitude of spin flipping depending on the orientation of neighboring spins.
This is directly connected to the fact that tunneling of particles depends on the occupation of the neighboring sites.
The constraint part of the Hamiltonian in the spin language then will forbid two next-nearest-neighboring spin-ups, i.e. the constraint is implemented only between spins of the same parity.

The second type of process gives the interaction part of the effective spin Hamiltonian
\begin{equation}
H_{U/2}^\mathrm{int}=-\frac{4}{15}\frac{J^2}{U/2}\sum_i(5+7\sigma_i^z+6\sigma_i^z\sigma_{i+1}^z+6\sigma_i^z\sigma_{i+3}^z),\label{eq:Heff_U2_int}
\end{equation}
where besides neighboring interactions, spins at distance three are coupled as well.

Adding together \eqref{eq:Heff_U2_tun} and \eqref{eq:Heff_U2_int}, and inverting $\sigma^x\to-\sigma^x$ to resemble the regime $E=U$ we obtain the effective Hamiltonian in its final form
\begin{gather}
H_{U/2}=\sum_i \Big[-\sqrt{2}\sigma_i^x+\lambda\sigma_i^\uparrow+W\sigma_i^\uparrow\sigma_{i+2}^\uparrow+\frac{8-56\sigma_i^\uparrow}{15}\nonumber \\
-2\sqrt{2}(\sigma_i^x\sigma_{i+1}^z+\sigma_i^z\sigma_{i+1}^x)-\frac{8}{5}(\sigma_i^z\sigma_{i+1}^z+\sigma_i^z\sigma_{i+3}^z)\Big].\label{eq:Heff_U2}
\end{gather}
where
\begin{equation}
\lambda=\frac{U/2-E}{J^2/(U/2)},
\end{equation}
denotes the detuning from the resonance in the regime
\begin{equation}
\begin{cases}
|U/2-E|\ll U\\
J\ll\mu\ll U
\end{cases},
\label{eq:tilt_superlat_regime}
\end{equation}
and the weight $W\to+\infty$ implements the constraint on spin configurations similarly to Eq.~\eqref{eq:Heff_U1}. Note that the Hamiltonian \eqref{eq:Heff_U2} is dimensionless as it is rescaled by a characteristic second order tunneling amplitude $J^2/(U/2)$.

Note that the first three terms are just two copies of Eq.~\eqref{eq:Heff_U1}, one for the even spins and one for the odd spins; these would be the only terms if the tunnelling of bosons did not depend on the occupation of neighboring sites.
The fifth and sixth terms represent the coupling between the odd and even sublattices that arises from this dependence (Fig.~1(c) in \cite{Buyskikh2018}).
The remaining term just shifts the entire energy spectrum along the energy and detuning $\lambda$ axes due to interactions between even and odd spins.

\section{Microscopic picture}
\label{sec:eff_model}

Essential points on the phase diagram for the spin models \eqref{eq:Heff_U1} and \eqref{eq:Heff_U2} become clear if one first takes a look at the cases of extremely large tilts, $\lambda\to\pm\infty$ (and $\tilde\lambda\to\pm\infty$).
Here we consider the infinite constraint case, $W=\infty$, i.e. forbidden spin configuration are completely removed from the Hilbert space.

Using a perturbative approach \cite{Bir1974} we determine the lowest excitations spectra and investigate similarities and differences between the models.
Both models have paramagnetic (PM) and antiferromagnetic (AFM) phases in the limit of large negative and positive tilts, respectively.
Each phase in both models has similar elementary excitations, however the interaction terms $\sigma^x\sigma^z$ and $\sigma^z\sigma^z$ in the regime $E=U/2$ \eqref{eq:Heff_U2} create new coupled excitations that become relevant at small field $\lambda$.

This replacement of the lowest excitations suggests that the nature of the phase transition changes as well.
In Sec.~\ref{sec:phase_transition} we investigate both models at their quantum critical points and confirm the prediction of this section --- different critical behaviors of the models.

\subsection{$E=U$ regime}

The analysis of this regime was partially presented in Ref.~\cite{Sachdev2002} and as it plays a role of a building block for the regime $E=U/2$ we present it here for the completeness of our description.
In the regimes of large tilts the spin flipping terms $\sigma^x$ in \eqref{eq:Heff_U1} are treated as perturbations.

\subsubsection{PM phase}

In the limit $\tilde\lambda\to+\infty$ the model is in the PM phase, where its ground state has all spins aligned along the longitudinal field, i.e. $\prod_i\ket{\downarrow}_i$.
In the language of bosons this corresponds to the state where all particles stay at the initial sites of the parent MI state.

The lowest elementary excitations are single spin-up states $\ket{j}=\ket{\uparrow}_j\prod_{i\neq j}\ket{\downarrow}_i$ (Fig.~\ref{fig:excitations}(c)), the degeneracy of which is lifted only in second order of the perturbation theory in $1/\tilde\lambda$ via the spin flipping terms $\sigma^x$.
In order for $\ket{j}$ to move the $j^\mathrm{th}$ spin should be flipped down and another spin flipped up, i.e. there are two possible channels depending on the order of these processes.
In general they cancel each other, however if the constraint for nearest-neighboring spins forbids one of the channels the excitation can move by one site.

In the thermodynamic limit, $M\to\infty$, using the effective Hamiltonian theory \cite{Bir1974} the lowest excited states energies above the ground state energy up to the second order read
\begin{equation}
\varepsilon_U^+(\tilde\lambda,k)=\tilde\lambda+\frac{8+4\cos(ka)}{\tilde\lambda}+\mathcal{O}\left(\tilde\lambda^{-2}\right),
\label{eq:eps_U_plus}
\end{equation}
where $k$ is the single excitation momentum and $a$ is the spacial separation between spins.

The higher excited states have two elementary excitations $\ket{j,j'}=\ket{\uparrow}_j\ket{\uparrow}_{j'}\prod_{i\neq j,j'}\ket{\downarrow}_i$, and its energy above the ground state energy scales as
\begin{equation}
\tilde{\varepsilon}_U^+(\tilde\lambda)=2\tilde\lambda+\mathcal{O}\left(\tilde\lambda^{-1}\right),
\label{eq:eps_tilte_U_plus}
\end{equation}
up to the first order corrections.
The interactions between excitations complicate the second order correction, but do not create any first order corrections, which will become essential in the regime $E=U/2$.

\subsubsection{AFM phase}

In the limit $\tilde\lambda\to-\infty$ the model is in the AFM phase, where its ground state maximizes the total number of spin-ups $\ket{\uparrow}$, and in order to obey the constraint spins are Neel ordered, i.e. $\ket{(\downarrow\uparrow)}$,
where $(...)$ implies periodicity.
In the language of bosons this corresponds to the state where only particles on every second site resonantly tunnel to their nearest-neighboring sites.

Note that for the systems of a finite size the degeneracy of the ground state depends on the type of boundary conditions as well as the parity of the number of spins $M$.
For instance, the ground state is twofold degenerate if $M$ is even and periodic boundary conditions (PBC) are imposed on the system, or the ground state is non-degenerate $M$ is odd and open boundary conditions (OBC).
We define the total number of spin-ups in the ground state as $M_\mathrm{gr}^\uparrow$, which in the thermodynamic limit makes the boundaries and parity irrelevant and approach $M/2$.

The lowest excited states have $M_\mathrm{gr}^\uparrow-1$ spin-ups and hence the Neel ordered phase should be broken somewhere.
Then a domain of Neel ordered spins is interrupted by a domain wall, which shifts one domain with respect to the other by one site.
Due to the constraint the domains can touch each other only via two consecutive spin-downs (Fig.~\ref{fig:excitations}(a)) and not two consecutive spin-up.
The state with $M_\mathrm{gr}^\uparrow-1$ spin-ups can have two domain walls and in thermodynamic limit they move independently, i.e. without interaction.

In the thermodynamic limit, $M\to\infty$, energies of states with a single domain wall with respect to the ground state energy up to the second order read
\begin{equation}
\varepsilon_U^-(\tilde\lambda,k)=|\tilde\lambda|+\frac{2-8\cos(2ka)}{|\tilde\lambda|}+\mathcal{O}\left(\tilde\lambda^{-2}\right),
\label{eq:eps_U_minus}
\end{equation}
where $k$ and $a$ are the single excitation momentum and spacial separation between spins.
Note that, comparing with \eqref{eq:eps_U_plus}, here the dispersion relation has periodicity two, which correctly reflects the order parameter (staggered magnetization) appearing in the AFM phase.

Analogous to the PM phase the higher excited states have two elementary excitations which energies above the ground state energy scale as
\begin{equation}
\tilde{\varepsilon}_U^-(\tilde\lambda)=2|\tilde\lambda|+\mathcal{O}\left(\tilde\lambda^{-1}\right),
\label{eq:eps_tilte_U_minus}
\end{equation}
up to the first order corrections.
The interactions between excitations complicate the second order correction, but the important part is that they do not create any first order corrections.

One should note that energies of the elementary excitations in both limits scale as $\sim\tilde\lambda$, Eqs.~\eqref{eq:eps_U_plus} and \eqref{eq:eps_U_minus}. 
Importantly, two elementary excitations do not couple together as their energies are affected only in the second order perturbation expansion, Eqs.~\eqref{eq:eps_tilte_U_plus} and \eqref{eq:eps_tilte_U_minus}.
This last point will become crucial for the regime $E=U/2$.

\subsection{$E=U/2$ regime}
\label{subsec:eff_model_U2}

The derivation of the perturbation Hamiltonian is similar to the regime $E=U$, the main difference now is that besides $\sigma^x$ perturbative terms there are also $\sigma^x\sigma^z$ and $\sigma^z\sigma^z$ terms which couple odd and even spins.
In the limit when perturbations are neglected completely the model is equivalent to two uncoupled spin chains of odd and even spins in the regime $E=U$ each.

However, a qualitative difference occurs in this model: coupling of elementary excitations lowers their mutual energy.
This creates coupled excitations --- new lowest energy excitations, which is the first indicator that the critical behavior of the model may be different from the regime $E=U$.
This prediction of microscopic consideration will be confirmed in Sec.~\ref{sec:phase_transition}.

\subsubsection{PM phase}

In the limit $\lambda\to+\infty$ the ground state of Eq.~\eqref{eq:Heff_U2} is a non-degenerate paramagnetic state $\prod_i\ket{\downarrow}_i$ and the lowest elementary excitations are single spin-up states $\ket{j}=\ket{\uparrow}_j\prod_{i\neq j}\ket{\downarrow}_i$ like in the regime $E=U$.

In the thermodynamic limit, $M\to\infty$, energies of the elementary excitations above the ground state read
\begin{equation}
\varepsilon_{U/2}^+(\lambda)=\lambda+\frac{64}{5}+\mathcal{O}\left(\lambda^{-1}\right),
\label{eq:eps_U2_plus}
\end{equation}
where the constant energy shift is acquired due to $\sigma^z\sigma^z$ terms in Eq.~\eqref{eq:Heff_U2}.
The degeneracy of these states is lifted only in the second order of the perturbation theory in $1/\lambda$ via simple spin flipping $\sigma^x$ and more complicated $\sigma^x\sigma^z$ terms.

Note that hopping of the elementary excitations in this case happens between spins of the same parity, i.e. second-neighbor interactions.
As a result the total spin chain in the regime $E=U/2$ can be treated as two spin chains $E=U$ , where elementary excitations move along each chain independently (Fig.~\ref{fig:excitations}(c)).

The higher excited states have two elementary excitations $\ket{j,j'}=\ket{\uparrow}_j\ket{\uparrow}_{j'}\prod_{i\neq j,j'}\ket{\downarrow}_i$ as well as in the regime $E=U$, however if these excitations are on different subchains they can couple together (Fig.~\ref{fig:excitations}(d)) via $\sigma^z\sigma^z$ terms and lower their mutual energy.
We will refer to them as coupled excitations with energies
\begin{equation}
\tilde{\varepsilon}_{U/2}^+(\lambda)=2\lambda+\frac{96}{5}+\mathcal{O}\left(\lambda^{-1}\right).
\label{eq:eps_tilde_U2_plus}
\end{equation}
One can observe that the energy of the coupled excitation is lower than the energy of two elementary excitations in isolation from one another.

\subsubsection{AFM phase}

In the limit $\lambda\to-\infty$ the ground state maximizes the total number of spin-ups in the system, but due to the constraint spins of the same parity are Neel ordered, i.e. $\ket{(\downarrow\downarrow\uparrow\uparrow)}$, where $(...)$ implies periodicity.

Note that the degeneracy of the ground state depends on the number of the spins $M$ and type of the boundary conditions, as well as it was in the regime $E=U$.
For instance, in the case of an even number of spins in each subchain and PBC the ground state is fourfold degenerate, in the case of an odd number of spins in both subchains and OBC the ground state is non-degenerate.
The total number of spin-ups in the ground state $M_\mathrm{gr}^\uparrow$ in the thermodynamic limit approaches $M/2$ as well as in the case $E=U$.

Similarly to the case $E=U$ the lowest excited states have $M_\mathrm{gr}^\uparrow-1$ spin-ups, which means that the Neel order of one of the subchains is broken by a domain wall, whereas the second chain is Neel ordered.
In the thermodynamic limit, $M\to\infty$, the lowest excited state energy above the ground state energy reads
\begin{equation}
\varepsilon_{U/2}^-(\lambda)=|\lambda|-\frac{16}{5}+\mathcal{O}\left(\lambda^{-1}\right),
\label{eq:eps_U2_minus}
\end{equation}
where the next order corrections include complicated contributions of spin flipping $\sigma^x$ and $\sigma^x\sigma^z$ terms, which are left outside the scope of this study.
The important part is the first order correction due to $\sigma^z\sigma^z$ terms which depend on the relative position of the domain wall with respect to the Neel ordered state of the other subchain.

The higher excited states have two elementary excitations and their mutual energy can be lowered by $\sigma^z\sigma^z$ terms.
We refer to them as coupled excitations as well with energy
\begin{equation}
\tilde{\varepsilon}_{U/2}^-(\lambda)=2|\lambda|-\frac{64}{5}+\mathcal{O}\left(\lambda^{-1}\right).
\label{eq:eps_tilde_U2_minus}
\end{equation}
One again observes that coupling of domain walls lowers their mutual energy and makes them energetically favorable at low values $\lambda$.

\subsection{Comparison}

\begin{figure}[tb]
\begin{centering}
\includegraphics[width=1\linewidth]{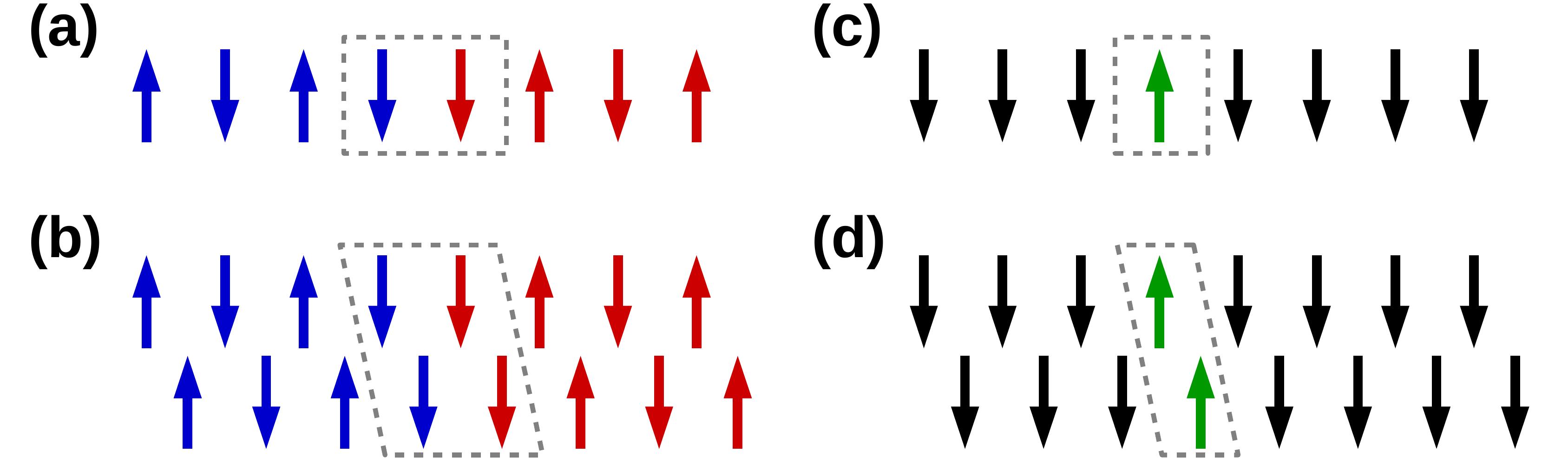}
\par\end{centering}
\caption{
\label{fig:excitations}
Illustrations of elementary (a,c) and coupled (b,d) excitations in the cases of extremely strong tilts $\tilde\lambda\to-\infty$ (a), $\lambda\to-\infty$ (b), $\tilde\lambda\to+\infty$ (c) and $\lambda\to+\infty$ (d).
The elementary excitations (a,c) are present in both regimes $E=U$ and $E=U/2$.
The coupled excitations (b,d) are present only in the regime $E=U/2$ where elementary excitations pair up via the interaction terms between even and odd spins.}
\end{figure}

From comparison of elementary excitations in both models in limits of large tilts $\lambda$ (or $\tilde\lambda$) we note an important difference.
In both regimes the lowest elementary excitations have the same nature (Fig.~\ref{fig:excitations}(a,c)).
In regime $E=U/2$ these elementary excitation are coupled and form pairs (Fig.~\ref{fig:excitations}(b,d)) with lowered mutual energy than in isolation from one another.
This process is impossible in the regime $E=U$.

This difference suggests that coupled excitations might become new lowest excitations as the model \eqref{eq:Heff_U2} approaches the phase transition at low values $\lambda$.
Of course, this perturbative consideration does not show the full picture, however, it reveals for instance the coupling role of $\sigma^z\sigma^z$ terms.

We will see that the prediction of this section is correct, we confirm it from calculations of the energy gap of Eq.~\eqref{eq:Heff_U2} in Fig.~1(a) of \cite{Buyskikh2018}.
One can see that the substitution of the lowest excitations takes place as the energy gap scaling changes from $\sim\lambda$ to $\sim2\lambda$.
In the next section we show that the critical behavior changes as well.

\section{Phase transition analysis}
\label{sec:phase_transition}

Traditionally, the existence in different regimes of the same model of an ordered and disordered phase is the smoking gun of the existence of a phase transition that separates the two.
The models described by the Hamiltonians \eqref{eq:Heff_U1} and \eqref{eq:Heff_U2} describe in different regimes AFM and PM phases and we thus expect that the two phases are separated by phase transitions.
In order to confirm our expectation, we perform a finite size scaling analysis of the order parameter.
In this way we can unveil the actual presence of a phase transition and locate the critical point $\lambda_\mathrm{crit}$.

First, we revisit the case $E=U$.
The existence of a phase transition in the Ising universality class has already been pointed out in \cite{Sachdev2002}.
Here we complete the identification by relating the model to the antiferromagnetic Ising chain in skew field (AFISF) \cite{Ovchinnikov2003}, which is known to host a second order phase transition belonging to the Ising universality class.

We extend the analysis to the regime $E=U/2$, where we successfully locate a new critical point. Our detailed finite-size scaling analysis suggests that the transition is in the tricritical Ising universality class.

\subsection{$E=U$ regime}

In this section we summarize the main aspects of the model \eqref{eq:Heff_U1} along with its symmetry content, extending the previous analysis in Ref.~\cite{Sachdev2002}.
By rearranging terms of the Hamiltonian it takes a more familiar form
\begin{equation}
\bar{H}_U=\sum_i\big(\sigma_i^z\sigma_{i+1}^z-h_x\sigma_i^x+h_z\sigma_i^z\big),\label{eq:Heff_U1_AFISF}
\end{equation}
where $h_x=4\sqrt{2}/W\to0$ is the amplitude of the transverse field and $h_z=2\left(\tilde\lambda/W+1\right)\to2$ the longitudinal field. The model \eqref{eq:Heff_U1_AFISF} is referred in literature as the AFISF model.

The previous numerical study of this model investigated the phase diagram in great detail \cite{Ovchinnikov2003}.
It can be mapped according to the behavior of the order parameter operator
\begin{equation}
{\cal M}_U=\sum_i(-1)^{i}\sigma_i^z,\label{eq:order_U1}
\end{equation}
representing the staggered magnetization of the spin chain.
The Neel ordered AFM phase in case of weak fields has $\braket{{\cal M}_U}\ne0$, in case of strong field the system is in the PM phase with $\braket{{\cal M}_U}=0$.
When the transverse field $h_x$ is exactly zero, the model becomes classical and the phase transition between the two phases is of first order. Otherwise the two phases are separated by a line of second order phase transitions of the Ising universality class.

\begin{figure}[tb]
\begin{centering}
\includegraphics[width=1\linewidth]{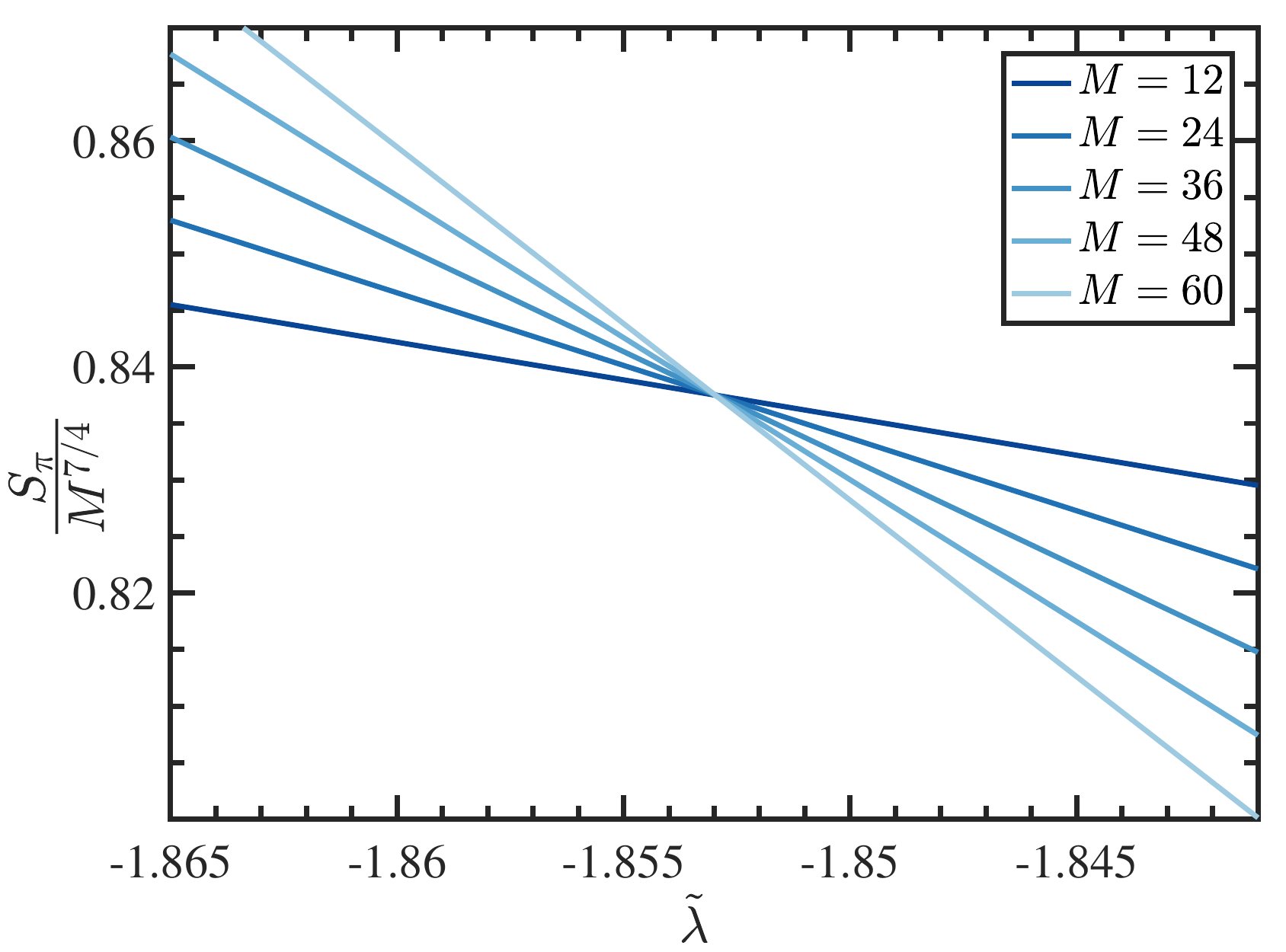}
\par\end{centering}
\caption{
\label{fig:str_fact_U1}
Scaling plot of the structure factor $S_{\pi}$ near the QCP in the regime $E=U$ for chains of $M$ spins with PBC.
According to the standard scaling argument \cite{Ginsparg1988} it scales as $\sim M^{2-\eta}=M^{7/4}$ for the Ising exponent $\eta=1/4$.
Calculations of eigenstates were performed using DMRG techniques and converged with the MPS bond dimension $D=96$.}
\end{figure}

The nature of the quantum critical point of second order is tested, for instance, by confirming numerically the anomalous scaling dimension of the order parameter $\eta$ that for the Ising universality class should be $\eta=1/4$.
We focus of the scaling of the structure factor
\begin{equation}
S_{\pi}=\sum_{i,j}(-1)^{i+j}\braket{\sigma_i^z\sigma_j^z},\label{eq:str_fact_U}
\end{equation}
which according to the standard scaling argument \cite{Ginsparg1988} scales as $\sim M^{2-\eta}$ at the critical point $\lambda_\mathrm{crit}$.
Using this we locate the critical point $\tilde\lambda_\mathrm{crit}\approx-1.853$ (Fig.~\ref{fig:str_fact_U1}), which is close to the asymptotic prediction (Eq.~(70) in \cite{Ovchinnikov2003}) $\tilde\lambda_\mathrm{crit}^\mathrm{asym}=-4\sqrt{2}/3\approx-1.886$.

Having confirmed the results of \cite{Sachdev2002} about the nature of the phase transition, we now proceed further and identify the $\mathcal{Z}_2$ symmetry that characterizes the Ising universality class and its breaking pattern.
Consider the Hamiltonian \eqref{eq:Heff_U1_AFISF} in the absence of the longitudinal field $h_z$, it is the standard transverse Ising model.
In this specific case it is easy to identify the symmetry operator as
\begin{equation}
G_\mathrm{TI}=\prod_i\sigma_i^x,\label{eq:symm_TI}
\end{equation}
that commutes with $\bar{H}_U$.
At the phase transition, the ground state of the system spontaneously breaks the $\mathcal{Z}_2$ symmetry, which means that in the AFM phase the ground state is two-fold degenerate and the symmetry operator $G_\mathrm{TI}$ maps one ground state into another inverting the sign of the order parameter $\braket{{\cal M}_U}$.
On the other side of the phase transition, in the PM phase, the ground state is unique and invariant with respect to the symmetry operator.

If $h_z\neq0$ the operator $G_\mathrm{TI}$ does not commute with the Hamiltonian anymore, i.e. $[G_\mathrm{TI},\bar{H}_U]\neq0$.
However the system still can undergo a phase transition associated with the $\mathcal{Z}_2$ symmetry breaking, hence there should be another symmetry breaking operator $G$ that commutes with the Hamiltonian $[G,\bar{H}_U]=0$ and squares to identity $G^2=\mathcal{I}$.
One can define $G$ as a single site translation operator
\begin{equation}
G[\sigma_i]=\sigma_{i+1},
\end{equation}
which shifts the entire spin chain by one site.
Another way to define $G$ is as an operator that inverts the entire spin chain of the length $M$ around the middle of a bond between two neighboring spins:
\begin{equation}
\begin{cases}
G[\sigma_i]=\sigma_{M-i}, & \mod(M,2)=0,\\
G[\sigma_i]=\sigma_{M-i+1}, & \mod(M,2)=1.
\end{cases}
\end{equation}
In the infinite chain limit, $M\to\infty$, boundary effects become irrelevant as well the position of the inversion center, thus we get $G^2=\mathcal{I}$ for both cases.

The proposed above symmetry operators $G$ exchange even and odd spins, hence the two Neel ordered ground states in the AFM phase are exchanged as well, whereas the PM state is left unchanged.
Thus $G$ is the operator that is spontaneously broken at the line of second order phase transitions of the model \eqref{eq:Heff_U1_AFISF}.

\subsection{$E=U/2$ regime}

In Ref.~\cite{Buyskikh2018} we have presented some results about the identification of the phase transition and its nature.
In particular, in Fig.~2(a) in \cite{Buyskikh2018} we locate the phase transition point as the place where the energy gap closes.
Moreover one can clearly see that the scaling of the gap changes from $\sim\lambda$ to $\sim2\lambda$ near the phase transition, which confirms the prediction of the microscopic consideration from Sec.~\ref{subsec:eff_model_U2} that coupled excitations dominate elementary excitations near the phase transition.

The scaling of the energy gap can be used to determine the exact location of the critical point $\lambda_\mathrm{crit}$ as well as the dynamical and correlation length critical exponents, but requires calculations of the first two eigenstates of the model with a high precision.
Here we focus on alternative methods using only the information obtained from the ground state.

Assuming that the critical point is conformally invariant we can use the results about the scaling of the entanglement entropy in conformal field theories in order to locate and characterize the critical point and predict scaling \cite{Callan1994,Osborne2002,Vidal2003,Calabrese2004,Calabrese2009}.
Using ideas similar to those proposed in the context of the phenomenological renormalization group \cite{Nightingale1975,Nightingale1982,Roncaglia2015,Roncaglia2008}, and first suggested in \cite{Koffel2012}, we determine the critical point $\lambda_\mathrm{crit}\approx-6.6676(1)$.
Scaling of the entanglement entropy at this point \cite{Buyskikh2018} is compatible with scaling of the tricritical Ising model with the central charge $c=7/10$ \cite{Guida1998}.

The term tricritical point was first used in discussions of the two-fluid critical mixing point in $\mathrm{He}_3-\mathrm{He}_4$ mixtures \cite{Griffiths1970} and since then attracted a lot of attention both theoretically and experimentally, which comprehensive reviews can be found in Refs.~\cite{Domb1984,Cardy2002}.
A tricritical point is defined as the end point of a line where three distinct phases coexist simultaneously, as contrasted with a critical point --- the end point of a line where two distinct phases coexist simultaneously.

A number of theoretical models have been confirmed to have tricritical points, for example Ising model with both ferro and antiferromagnetic interactions (metamagnets model) \cite{Nelson1975}, Blume-Emery-Griffiths model \cite{Blume1971}, Potts model \cite{Potts1952}, Ashkin-Teller model \cite{Ashkin1943}, and more general $n-$state cubic model \cite{Kim1975}.
An example of the model with the particular tricritical Ising point is the Ising model with vacancies \cite{Cardy2002}, which is a generalization of the Ising model where spins can be absent.

One of the main features of a tricritical point is the existence of two relevant symmetry breaking operators that need to be simultaneously tuned to their critical value in order to observe the desired critical behavior.

Here we support the results presented in \cite{Buyskikh2018} about the presence of a tricritical Ising point in the regime $E=U/2$ by analyzing the finite size scaling behavior of the order parameter
\begin{equation}
{\cal M}_{U/2}=\sum_i \mathrm{e}^{\mathrm{i} \frac{\pi}{2} i} \sigma_i^z,\label{eq:order_U2}
\end{equation}
the staggered magnetization with a period of four spins.
This choice of the order parameter operator is based on the microscopic consideration in Sec.~\ref{subsec:eff_model_U2}.
In the disordered PM phase spins are aligned with the external field, i.e. $\braket{{\cal M}_{U/2}}=0$, in the ordered AFM state spins of the same parity are Neel ordered, i.e. $\braket{{\cal M}_{U/2}} \ne 0$.
The symmetry breaking at the critical point $\lambda_\mathrm{crit}$ implies that one out of four possible states is chosen in the AFM phase.

In order to test scaling of the order parameter we compute a real part of the structure factor
\begin{equation}
S_{\pi/2}=\sum_{i,j}(-1)^{i+j}\braket{\sigma_{2i}^z\sigma_{2j}^z+\sigma_{2i-1}^z\sigma_{2j-1}^z}.\label{eq:str_fact_U2}
\end{equation}
The two symmetry breaking operators have scaling dimensions $\eta=3/20$ (primary field) and $\tilde{\eta}=7/4$ (subleading field) \cite{Guida1998,Henkel1999,Francesco1997,Cardy2002}.
Hence the structure factor $S_{\pi/2}$ scaling at the phase transition point $\lambda_\mathrm{crit}$ should have two contributions as well, in particular $\sim M^{2-\eta}$ and $\sim M^{2-\tilde{\eta}}$.
The existence of two symmetry breaking operators at the tricritical point makes its scaling analysis substantially harder than in cases of a critical point with a single symmetry breaking operator.
In particular only in the limit of large system sizes $M$ the primary contribution dominates the scaling and the lack of perfect collapse should be attributed to the presence of the the subleading operator.

\begin{figure}[tb]
\begin{centering}
\includegraphics[width=1\linewidth]{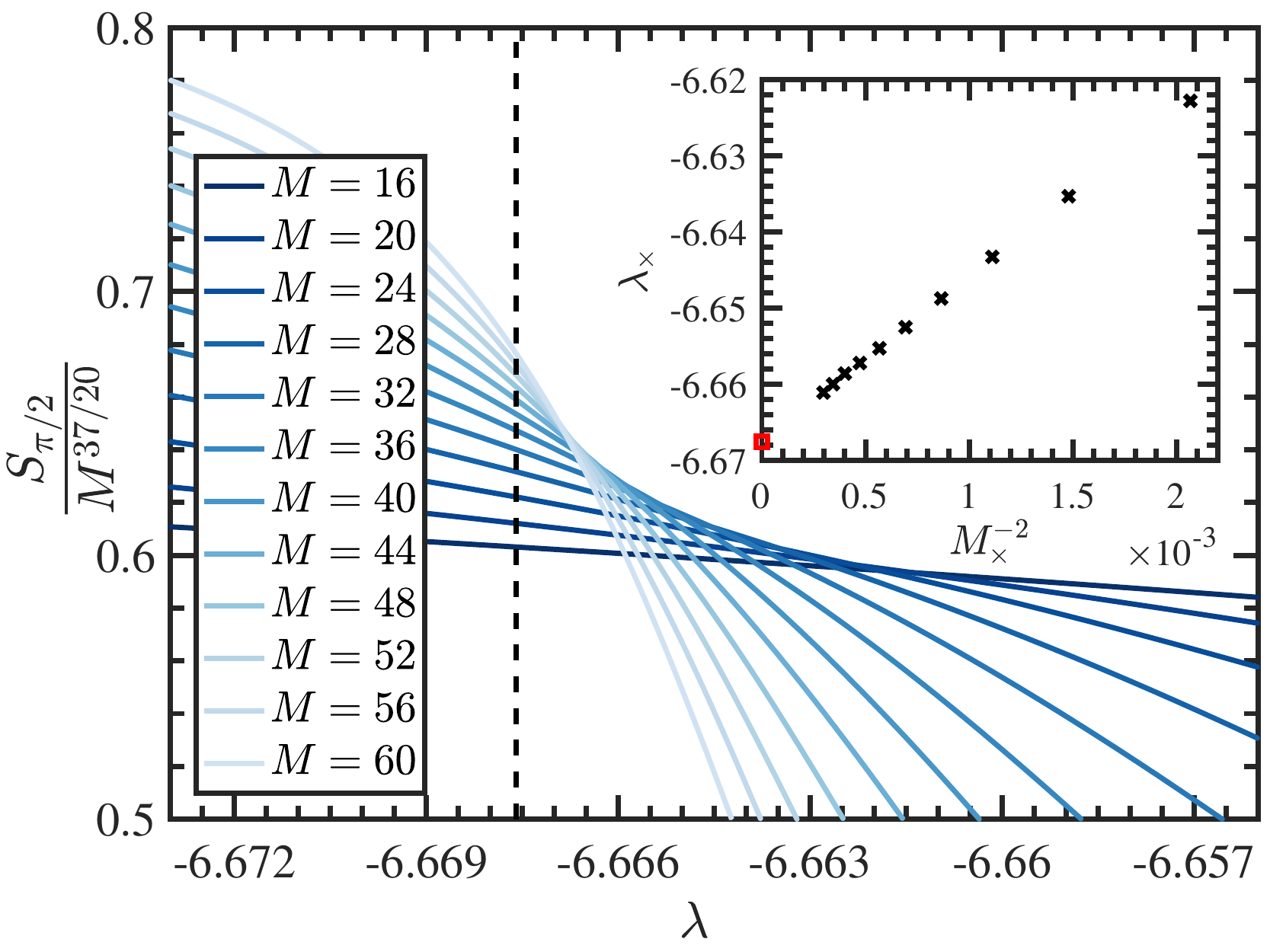}
\par\end{centering}
\caption{
\label{fig:str_fact_U2}
Scaling plot of the structure factor $S_{\pi/2}$ near the QCP in the regime $E=U/2$ for chains of $M$ spins with PBC.
Scaling is preformed for the leading magnetization exponent $\eta=3/20$ of the tricritical Ising point.
The convergence is much slower that in the case of the Ising critical point (Fig.~\ref{fig:str_fact_U1}).
Here the convergence is slowed down by a subleading magnetization field corrections $\sim M^{2-\tilde{\eta}}$, where $\tilde{\eta}=7/4$ for the tricritical Ising point.
The dashed line denotes $\lambda_\mathrm{crit}=-6.6676(1)$ found via scaling of the von Neumann entropy in \cite{Buyskikh2018}.
The inset shows crossings of the Binder cumulants $U_M$ (see the main text) converging to the same critical point $\lambda_\mathrm{crit}$ (denoted as a red square).
Calculations were performed using DMRG techniques and converged with the MPS bond dimension $D=512$.}
\end{figure}

In Fig.~\ref{fig:str_fact_U2} we present the structure factor $S_{\pi/2}$ scaling.
As the system size $M$ increases we see a slow convergence toward $\lambda_\mathrm{crit}$ (dashed line), which shows compatibility with the tricritical Ising scenario.

Our last test of the phase transition is based on the Binder cumulant \cite{Binder1981}, which allows us to define $\lambda_\mathrm{crit}$ without a prior knowledge of the universality and critical exponents.
For that we calculate
\begin{equation}
U_M=3\Braket{(\Delta{\cal M}_{U/2})^2}_M^2-\Braket{(\Delta{\cal M}_{U/2})^4}_M,\label{eq:binder_cumulant}
\end{equation}
where $\Delta{\cal M}_{U/2}={\cal M}_{U/2}-\braket{{\cal M}_{U/2}}_M$ and the index $M$ indicates the length of the spin chain.
At the critical point $U_M$ approaches a constant value.

The method works in the following way: for each pair of lines $U_M$ and $U_{M+4}$ we find the intersection point $\lambda_\times$ and define the corresponding average system size as $M_\times=M+2$.
In the inset of Fig.~\ref{fig:str_fact_U2} one can see the convergence of $\lambda_\times (M_\times)$ towards $\lambda_\mathrm{crit}$ (red square) found form the von Neumann entanglement scaling \cite{Buyskikh2018}.
This shows a good agreement between both methods, and that the method using the entanglement scaling allows us to define $\lambda_\mathrm{crit}$ with a higher precision.

Finally we note that without $\sigma^x\sigma^z$ and $\sigma^z\sigma^z$ interactions in Eq.~\eqref{eq:Heff_U2} the critical behavior is completely identical to the case of decoupled $E=U$ chains with $\mathcal{Z}_2\otimes\mathcal{Z}_2$ symmetry, i.e. two copies of the standard Ising transition.
However, the presence of the interactions $\sigma^x\sigma^z$ and $\sigma^z\sigma^z$ changes the criticality of the model to the tricritical Ising phase transition, which is associated with a $\mathcal{Z}_2$ symmetry breaking.
Due to the complicated form of the Hamiltonian \eqref{eq:Heff_U2} we have not succeeded in identifying a corresponding symmetry operator and leave this question open.

\section{Finite temperature behavior}
\label{sec:finite_temp}

In the following section we compare the behavior of models \eqref{eq:Heff_U1} and \eqref{eq:Heff_U2} at finite temperatures.
A quantum system near its criticality is especially sensitive to thermal fluctuations because of rapid closure of the energy gap.
We investigate the specific heat capacity per spin
\begin{equation}
c(T,\lambda)=\frac{1}{T^2 M}\Braket{\Delta H^2(\lambda)}_T,
\end{equation}
which is a dimensionless measure of heat transfer between eigenstates of the system. Here $\braket{\Delta H^2(\lambda)}_T$ is the variance of the total energy in the system at temperature $T$, which is also a dimensionless parameter since the Hamiltonians \eqref{eq:Heff_U1} and \eqref{eq:Heff_U2} are made dimensionless by appropriate rescalings.

\begin{figure}[tb]
\begin{centering}
\includegraphics[width=1\linewidth]{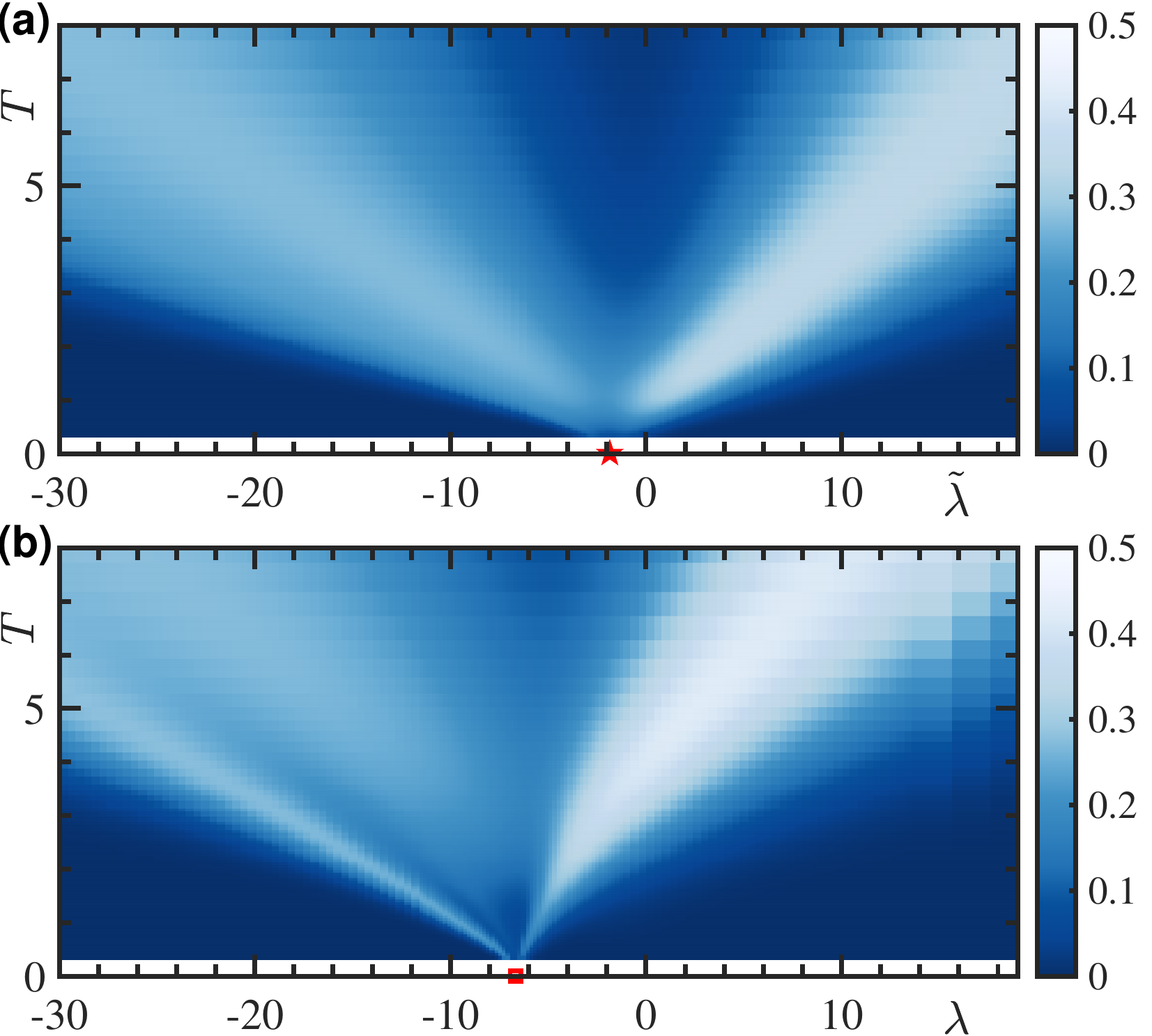}
\par\end{centering}
\caption{
\label{fig:Cv_U1_2}
Specific heat capacity per spin $c(T,\lambda)$ near quantum critical points in regime (a) $E=U$ for $M=101$ spins (b) $E=U/2$ for $M=102$ spins, both with OBC.
(a) We can see convergence of the specific heat peaks to the QCP $\tilde\lambda_\mathrm{crit}=-1.853$, at $T=0$ (red star).
(b) The complicated structure of the specific heat reflects the energy gap diagram in Fig.~2(a) of \cite{Buyskikh2018}, where one can see the suppressed transition of elementary excitations and the transition of coupled excitations --- true quantum phase transition.
There are two branches of specific heat peak diverging from \textit{each} phase transition point.
Presumably, two right branches merge together and cannot be distinguished, that is why one can observe only three branches.
Two of those branches converge to the critical point of $E=U/2$ model, $\lambda_\mathrm{crit}=-6.6676(1)$ (red square).
The third and forth branches correspond to the transition of elementary excitations and vanish at finite temperatures as the corresponding transition is suppressed and the energy gap minimum stays finite.
The results are obtained via TDVP evolution of the infinite temperature density matrix with bond dimension $D=128$.}
\end{figure}

We find that behavior of $c(T,\lambda)$ at critical points in regimes $E=U$ and $E=U/2$ have clear distinctions that are compatible with investigations of the lowest excitations in the system.
The regime $E=U$ has a simple Ising critical point and addition of a finite temperature $T$ to the system washes away its position due to thermal fluctuation (Fig.~\ref{fig:Cv_U1_2}(a)).
One can see two branches of $c(T,\lambda)$ diverging from the QCP as $T$ increases.

In the regime $E=U/2$ the nature of the transition seems to be more complicated (Fig.~$\ref{fig:Cv_U1_2}$(b)).
The specific heat dependence changes on both sides of the transition, i.e. it is not as symmetric as in regime the $E=U$. One can also note a change in the curvature of branches diverging from the QCP.

Another significant difference in the regime $E=U/2$ is the appearance of the third branch of $c(T,\lambda)$.
It can be explained via the energy gap diagram from Fig.~2(a) of \cite{Buyskikh2018}, where the suppressed transition of elementary excitations is still visible on the plot along with the true quantum phase transition of coupled excitations.

Each phase transition point has two branches of specific heat peaks diverging from it as the temperature increases.
We believe that Fig.~\ref{fig:Cv_U1_2}(b) has four branches in total, but two right branches merge together and cannot be distinguished.
Among those four, two branches converge to the critical point of the $E=U/2$ model and two other vanish at a finite temperature as the corresponding suppressed transition at $\lambda\approx-1.853$ always has a finite energy gap.

\section{Summary and outlook}
\label{sec:summary}

In this work we derived and investigated effective spin models for the unit filled Bose-Hubbard model quenched to regimes with a linear tilt enabling resonant transitions between nearest-neighboring and next-nearest-neighboring sites.
For the latter the superlattice geometry is required to map the motion of atoms to effective spin chains.

The first model \eqref{eq:Heff_U1} is derived for tilt values near $E=U$, i.e. bosons resonantly tunnel to neighboring sites.
This model is equivalent to the antiferromagnetic Ising chain in skew field (AFISF) with infinite, projective-like interactions.
Ordered and disordered phases in this model are separated by a second order phase transition, which belongs to the Ising universality class.
We identify a corresponding symmetry operator $G$ that breaks the $\mathcal{Z}_2$ symmetry as the system undergoes the phase transition.

The second model \eqref{eq:Heff_U2} is derived for tilt values near $E=U/2$ and superlattice geometry $U\gg\mu\gg J$, i.e. bosons can resonantly tunnel only to next-nearest-neighboring sites.
This spin chain is equivalent to a pair of spin chains in the regime $E=U$ coupled to each other via $\sigma^x\sigma^z$ and $\sigma^z\sigma^z$ interactions.
Analogous to the first model, a quantum critical point also separates ordered and disordered phases, but the nature of the phase transition is different.
The finite size scaling shows that the critical point belongs to the tricritical Ising universality class, which is associated with the $\mathcal{Z}_2$ symmetry breaking.
The exact form of the symmetry operator in this case stays unresolved though.

Besides universality classes of the phase transition the underlying differences between the two models can be seen from the analysis of effective models for extreme values of the field $\lambda$ (or $\tilde\lambda$).
Without interaction terms the $E=U/2$ model will be equivalent to two decoupled $E=U$ chains with independent spectrum of elementary excitations in each chain.
However, the $\sigma^z\sigma^z$ interactions couple these elementary excitations lowering their mutual energy.
As a result the role of elementary excitations at the QCP is suppressed by coupled excitations, which we directly observe in the energy gap diagram in Fig.~2(a) of \cite{Buyskikh2018}.

The presented result can be easily extended to the case of not only unit filling, but any uniform integer filling.
The main difference in spin models then will be due to the Bose enhancement factor of tunneling and a number of two-body interactions at each sity.
Also the superlattice offset $\mu$ is a free parameter of the model and hence can be exploited.
In the generic case of $J\ll\mu\lesssim U$ matrix elements for even and odd spins of $E=U/2$ model will be different and depend on $\mu$ as well, then Eq.~\eqref{eq:Heff_U2} will be obtained in the limit $\mu/U\to0$.
The study of $\mu-$dependence has a potential interest as it can allow us to modify the phase transition universality class, however this question is left outside of this paper.

Currently, a number of experimental groups have already performed experimental investigations of bosonic species in tilted optical lattices near resonances $E=U/n$, with integer $n$.
In this regimes the system dynamics truly obey quantum many-body physics with minimal influence of the environment, which was the main motivation for this paper.
Now with deeper theoretical understanding of the Bose-Hubbard model behavior near resonances, via effective models, it will be even more interesting to perform experiments with bosons in tilted optical lattice, but with superlattice geometry.
Another way to realize the Bose-Hubbard Hamiltonian \eqref{eq:H_BH}, but without the external linear field, is to use time-dependent tunneling amplitudes $J\mathrm{e}^{-i\omega t}$ in one direction of hopping and $J\mathrm{e}^{+i\omega t}$ in the other direction instead of time-independent $J$, here $\omega$ is equivalent to the linear field $E$.

The data for this manuscript is available in open access at Ref.~\cite{data_long}.

\begin{acknowledgements}

We thank Pasquale Calabrese, Florian Meinert, Manfred Mark, Giuseppe Mussardo, Roger Mong, Hanns-Christoph N\"agerl, Subir Sachdev, and Jon Simon for helpful and simulating discussions. Work at the University of Strathclyde was supported by the EPSRC Programme Grant DesOEQ (EP/P009565/1), by the European Union Horizon 2020 collaborative project QuProCS - Quantum Probes for Complex Systems (grant agreement 641277), and by the EOARD via AFOSR grant number FA2386-14-1-5003. D.S.~acknowledges support of the D-ITP consortium, a program of the Netherlands Organisation for Scientific Research (NWO) that is funded by the Dutch Ministry of Education, Culture, and Science (OCW). D.P.~was supported by the Charles E Kaufman foundation. Results were obtained using the EPSRC funded ARCHIE-WeSt High Performance Computer (EP/K000586/1).

\end{acknowledgements}

\appendix

\section{Numerical methods}
\label{sec:app_numerics}

Our analysis of the critical behavior of the spin models heavily relies on the numerical calculations of the eigenstates and thermal states.
The constraints on spin configurations in models \eqref{eq:Heff_U1} and \eqref{eq:Heff_U2} allows us to significantly reduce the Hilbert space size.
Following this trick one can perform ED calculations with almost 30 spins on a regular computer.

In order to go beyond one should use DMRG techniques with tensor networks of matrix product states and operators (MPS/MPO).
The constraint terms in Eqs.~\eqref{eq:Heff_U1} and \eqref{eq:Heff_U2} can be implemented via taking a limit of large $W$, which will not necessary keep numerical calculations stable and instead add another numeral parameter.
Then the implementation of the constraints can be realized via symmetrical tensors analogous to Refs.~\cite{Singh2011,Singh2012}.

Here we present an alternative way of implementing the spin-restricting constraints via projectors on the reduced Hilbert space.
The main advantage of this method is that it is compatible with any non-symmetric tensor network code and can be generalized for more complicated symmetries and constraints.

We first present the general form projectors and then give an example how these projectors can be embedded into already existing methods on example of the thermal state calculation.
Using this approach we obtain the system state time-evolution and eigenstates at QCP for $M=300$ spins (OBC) and $M=60$ spins (PBC) with convergence in the MPS bond dimension $D=512$.

\subsection{Realization of constraints via projectors}

All forbidden spin states can be projected out from the Hilbert space using the following operator
\begin{equation}
P_{U/n}=\prod_i(\mathcal{I}-\sigma_i^\uparrow\sigma_{i+n}^\uparrow),\label{eq:proj}
\end{equation}
which explicitly forbids states with a pair of spin-ups at distances $n=1$ or 2, for models in the regime $E=U$ and $E=U/2$, respectively.
In case of PBC imposed on the system, cyclic conditions are used for spin indices.

\begin{figure}[tb]
\begin{centering}
\includegraphics[width=1\linewidth]{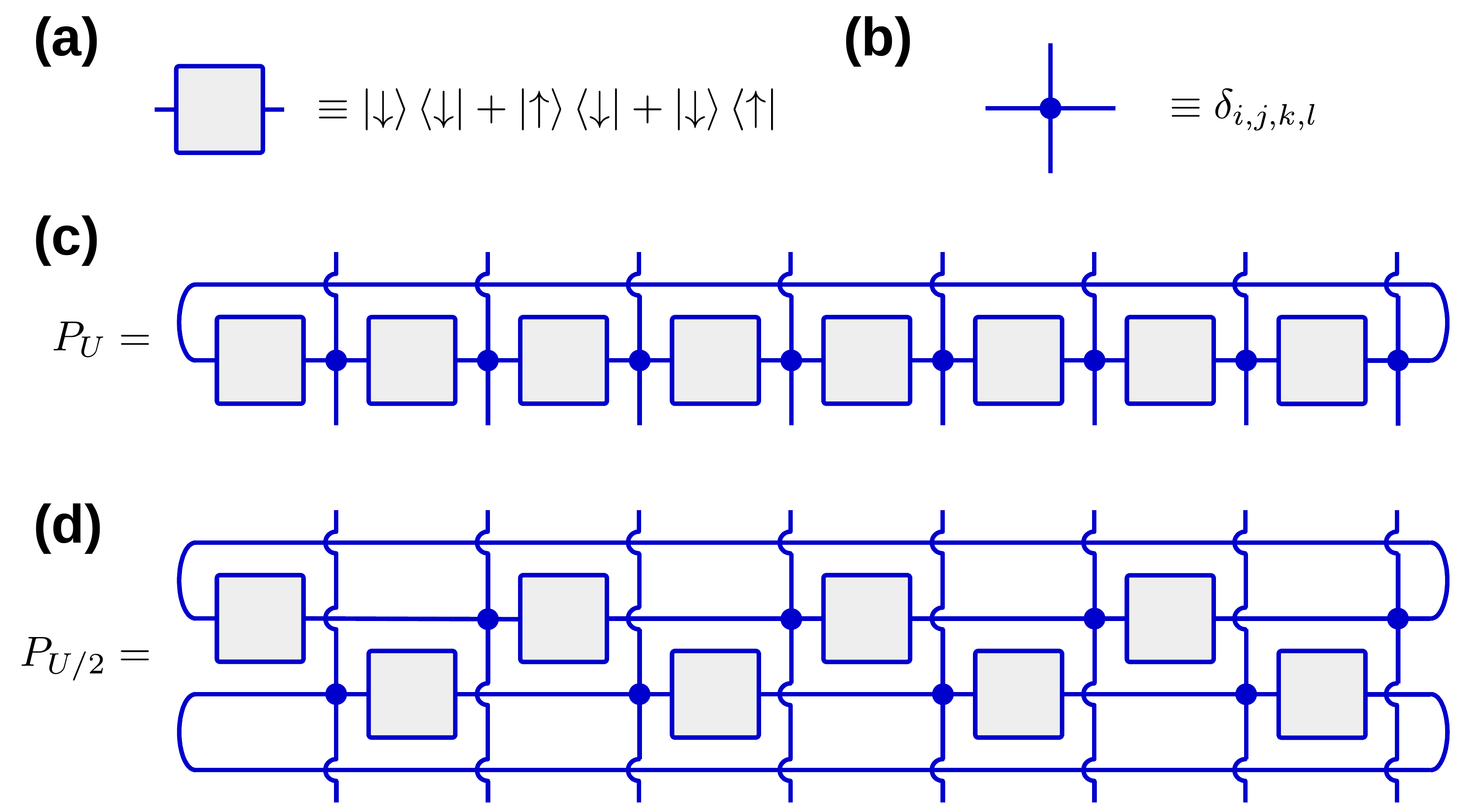}
\par\end{centering}
\caption{
\label{fig:projectors}
Examples of the spin-restricting projectors $P_{U/n}$ in the MPO form.
(a) Transfer matrix realizes the constraint on two connected spins.
(b) Rank-4 copy tensor.
(c) Projector $P_U$ implements the constraint in the regime $E=U$ with PBC.
In case of OBC one needs to remove contraction of the first and last tensors and the corresponding transfer matrix.
(d) Projector $P_{U/2}$ implementing the constraint for the regime $E=U/2$ with PBC and an even number of spins.}
\end{figure}

These projectors have a compact MPO representation via a network of transfer matrices and copy tensors (Fig.~\ref{fig:projectors}).
Note that in case of an odd number of spins, the form of $P_{U/2}$ will be slightly different.

\subsection{Example: T-MPS calculations}

Then Hamiltonians $H_{U/n}$ (Eqs.~\eqref{eq:Heff_U1} and \eqref{eq:Heff_U2}) are replaced by
\begin{equation}
\tilde{H}_{U/n}=P_{U/n} H_{U/n} P_{U/n},\label{eq:PHP_Un}
\end{equation}
where the constraint terms disappear naturally. These new composite Hamiltonians can be used for all sorts of calculations in the restricted Hilbert space, both in statics and dynamics.

For instance, we obtain the imaginary-time evolution of the initial infinite temperature density matrix to finite temperatures by means of the TDVP algorithm \cite{Haegeman2011,Koffel2012,Haegeman2013,Haegeman2016}.
In the Hilbert space without any restrictions the infinite-$T$ density matrix is proportional to identity, $\rho_0\propto I$. However, in the presence of restrictions one must take into account only allowed states, such that $\rho_0\propto P_{U/n}\mathcal{I}P_{U/n}=P_{U/n}$.

The next step is to evolve the density matrix to finite temperatures $\rho(\beta)\propto\mathrm{e}^{-\beta\tilde{H}}$, where $\beta=1/T$.
In order to preserve positive semi-definiteness of the density matrix we used the purification technique \cite{Verstraete2004,Cuevas2013} and rewrite this expression as $\rho(\beta)\propto\mathrm{e}^{-\beta\tilde{H}/2}\rho_0\mathrm{e}^{-\beta\tilde{H}/2}$.
In this case only one side of the density matrix is evolved $\bar{\rho}(\beta)\equiv\mathrm{e}^{-\beta\tilde{H}/2}\rho_{0}$, and since $\rho_0^2=\rho_0=\rho_0^\dagger$ expectation values can be obtained as
\begin{equation}
\braket{\hat{O}}_\beta=\frac{\mathrm{tr}[\hat{O}\bar{\rho}(\beta)\bar{\rho}^\dagger(\beta)]}{\mathrm{tr}[\bar{\rho}(\beta)\bar{\rho}^\dagger(\beta)]},
\end{equation}
where $\hat{O}$ is an arbitrary operator in MPO form.

Furthermore, one can reduce the computation cost and compress bond dimensions of sparse $\tilde{H}_{U/n}$ in MPO form by performing compression procedure developed for compression of MPS (Sec.~2.2 in \cite{Garcia-Ripoll2006}) via variational minimization of the distance between states.
For instance, in the case of open boundary conditions the original bond dimensions of $P_{U/2}$ and $H_{U/2}$ are 4 and 6, respectively, if they are constructed in a sparse way.
The resulting bond dimension of $\tilde{H}_{U/n}$ is $D_\mathrm{MPO}=4\times6\times4=96$, but it can be compressed down to just $D_\mathrm{MPO}=8$ with desired machine precision that does not exceed the major numerical error --- truncation error.

\bibliographystyle{apsrev4-1}
\bibliography{references}

\end{document}